# Dielectric properties of Granodiorite partially saturated with water and its correlation to the detection of seismic electric signals

A.N. Papathanassiou *, I. Sakellis *, and J. Grammatikakis

University of Athens, Physics Department, Section of Solid State Physics, Panepistimiopolis, 15784 Zografos, Athens, Greece

## ABSTRACT

Transient electric signals emitted prior to earthquake occurrence are recorded at certain sites in the Earth's crust termed sensitive. These field observations enforce the laboratory investigation of the dielectric response of rocks forming these localities. The dielectric relaxation of granodiorite rock coming from such a sensitive locality (Keratea, Greece) reveals, through complex impedance spectroscopy, that the activation volume for relaxation of this rock is negative which so far has been reported only rarely. This result, however, supports a theoretical model on the pre-seismic electric signals and is likely to be correlated with the sensitivity of the site and hence with the selectivity

*Corresponding authors; E-mail addresses: antpapa@phys.uoa.gr (A.N.P); e_sakel@phys.uoa.gr (I.S.)

# 1. Introduction

Seismic electric signals (SES) are transient electric signals recorded before an earthquake occurrence (Varotsos and Alexopoulos, 1984, a,b). Signals are collected by stations (consisting of electrode arrays of certain arrangement and scale) situated at on to "sensitive localities" in the Earth's crust. In these sensitive localities, SES signals are considered to be enhanced. Stations located at sensitive positions collect SES stemming from restricted seismic areas which may be far from the station (selectivity effect). The latter can be related with both large-scale (or "regional") and small-scale (or "local") heterogeneous properties of the earth's crust (Varotsos and Lazaridou, 1991). Field observations and theoretical models indicate that in a strongly inhomogeneous area some sub-areas may provide localities, which may be sensitive to SES. One way to make a sites sensitive may be that one or both electrodes are located within a structure in which SES are significantly enhanced. A typical example of such a case is Keratea station, Greece (Varotsos and Lazaridou, 1991) . This station lies in a place with dykes of granodiorite, with surface width of the order of a few meters. Field observations showed that when one or both electrodes are installed on the granodiorite dyke, strong SES are observed (Varotsos and Lazaridou, 1991).

The sensitivity effect of a SES station might reasonably be related to some specific physical) characteristics of the rocks underlying the station. So, it was considered highly desirable to investigate the dielectric properties of granodiorite of Keratea station, to shed light on one of the parameters that probably constitute the sensitivity phenomenon, i.e., the role of the type of the sustrating rock that constitutes the site. In the present work, we study the dielectric relaxation properties of the granodiorite from the Keratea station, under different states of hydrostatic pressure and temperature in the laboratory by employing complex impedance measurements in the frequency domain. As it will be evident in this work, this granodiorite belongs to the class of solids that are characterized by negative activation (relaxation) volume, which is reported in the literature very rarely.

# 2. Effect of pressure on the dielectric relaxation

A consequence of the presence of aliovalent impurities in crystalline ionic solids is the formation of extrinsic defects, such as vacancies and interstitials, which are produced for reasons of charge compensation (e.g., see Varotsos (2007) and references therein). A portion of these vacancies and/or interstitials form electric dipoles with the aliovalent impurities (defect dipoles). The dynamics of these dipoles, when triggered by an external 'force' (e.g., an electric field), yield dielectric relaxation, that is determined (within the frame of the rate process theory) by the relaxation time:

$$\tau(P,T) = (\lambda \nu)^{-1} \exp(g^{act}/k_B T) \qquad (1)$$

where P and T denote the pressure and temperature, respectively, $\nu$ is the vibration frequency of the relaxing charge carrier, $\lambda$ is a geometrical constant, $g^{act}$ is the Gibbs activation energy for relaxation and $k_B$ is the Boltzmann's constant. Note that all thermodynamic quantities appearing in this paper are related to relaxation or, alternatively, to the localized motion of the electric charge species that form the relaxing 'dipoles'. By differentiating Eq. (1) with respect to pressure at constant temperature, we get:

$$(\partial \ln \tau / \partial P)_T = -\frac{\gamma}{B} + \frac{\upsilon^{act}}{k_B T} \qquad (2)$$

where $\gamma \equiv -(\partial \ln \nu / \partial \ln V)_T$ is the Grüneisen constant (V denotes the volume), $B \equiv -(\partial P / \partial \ln V)_T$ is the isothermal bulk modulus and

$$\upsilon^{act} \equiv (\partial g^{act} / \partial P)_T \qquad (3)$$

is the activation volume (Varotsos and Alexopoulos, (1986)).

Structural heterogeneity of rocks due to porosity and distribution in the size of grains or crystallites, combined with compositional heterogeneity stimulates interfacial polarization phenomena. Strong polarization phenomena appear when porosity is filled (partially or fully) with water (Endres and Knight, 1991; Nettelblad, 1996; Papathanassiou, 2000a, b , 2001a, b; Varotsos, 2005). Additionally, water in rocks (which is abundant in the earth's crust) plays a dual role: (i) it enhances the dielectric properties of the rock and (ii) induces the so-called *dilatancy* when the rock is compressed. In particular, the latter occurs in the hypocenter area of an earthquake which is surrounded by water saturated porous rock with fluid-filled pore channels and the pre-earthquake stage is assumed to be accompanied by the appearance of fresh cracks in the fracture zone (Surkov et al, 2002). The time variation of the stress field prior to an earthquake results in the dynamic compression of water-filled rocks (Morgan and Nur, 1986; Morgan et al, 1989). On account of these reports, we presume that the dielectric response of a rock results from different dynamic processes, such as defect dipole rotation, interfacial polarization, double layer polarization, etc, which can be regarded generally as effective electric 'dipoles'. These electric 'dipoles' in rocks, tend to orient themselves under the influence of polarizing fields such as the stress field, which plays a role similar to that of an external field (Varotsos 2005). , When they undergo a monotonic compression, their characteristic relaxation time is modified with increasing pressure. Eq. (2 1) indicates that the percentage variation of the relaxation time caused by an isothermal change in pressure is determined by $v^{act}$. A negative activation volume implies that the rotational mobility of electric 'dipoles' is enhanced on increasing pressure, i.e., the characteristic relaxation time becomes shorter on pressurization. Negative activation volumes were experimentally found earlier in a few materials, by studying the pressure dependence of the electrical conductivity ( e. g., in NAFION hydrogels (Fontanella et al, 1996)) or the pressure dependence of the dielectric relaxation (e, g., in $\beta$-$PbF_2$ doped with lanthanum (Fontanella et al, 1982), and in semi-conducting polypyrrole (Papathanassiou et al, 2006, 2007)). Recently, negative activation volumes were found in rocks filled with water, such as leukolite (polycrystalline $MgCO_3$) and kataclastic limestone (polycrystalline $CaCO_3$) (Papathanassiou et al (2010)). The latter finding for a natural rock is particularly important for the following reason: Varotsos, Alexopoulos and Nomicos (Varotsos et al, 1982, see also Varotsos and Alexopoulos 1986) in order to explain the generation of seismic electric signals (SES) (i.e., the electric signals that are observed (Varotsos, 2005; Varotsos at al, 2006a, b; Varotsos and Alexopoulos, 1984a, b; Varotsos et al, 2005; 2003a, b; Varotsos et al, 2002) before earthquakes, regarded the earth's crust as a solid containing 'dipoles' in a polarizing field (i.e., the mechanical stress field). The monotonic increase of pressure exerted on rocks characterized by negative activation volume for relaxation, during the earthquake preparation, reduces the relaxation time of these 'dipoles'. Hence, at a certain critical pressure (smaller than that at which rupture occurs) the relaxation time of 'dipoles' becomes too short and then the 'dipoles' undergo a transition from random-orientation state to an oriented one. The time variation of the polarization of the rock yields the emission of a transient polarization current, which is SES. This is known in solid state physics as pressure stimulated currents (PSC) (for a brief review of PSC see also Varotsos et al., 1998).

### 3. Experimental setup and methodology

Granodiorite rock samples from Keratea (Greece) of thickness ranging from 1 to 2 mm and parallel surfaces of 1-2 cm$^2$ were placed inside the pressure vessel, a Novocontrol (Germany) high pressure apparatus operating from ambient pressure to 0.3 GPa and at various temperatures ranging from room temperature to 393 K. The pressurization fluid was slicon oil. Metallic electrodes were pasted on to the parallel surfaces of the specimen and a very thin insulating layer of epoxy covered the specimen to prevent contamination from the pressure transmitting fluid (Papathanassiou and Grammatikakis, 1996; Papathanassiou, 1998, 2001a,b; 2002). Note that the use of epoxy layer to jacket the sample was also suggested independently by the manufacturer of the pressure apparatus. The pressure system was connected with an HP impedance analyzer operating from 10$^{-2}$ to 10$^6$ Hz. Measurements in the frequency domain were monitored by a computer. The rock samples were partially saturated with water by immersing (and keeping) them in distilled water at 373 K for a time duration of about 3 months. The elevated temperature yield expansion of the specimen and opening of the pore cavities and, at the same time, diffusion of water molecules into the rock samples is enhanced. By weighting the sample before and after the wetting we found that the (partially) saturated rock accommodates about 0.5 wt % water.

The tangent loss angle function $\tan\delta \equiv \text{Im}(\varepsilon)/\text{Re}(\varepsilon)$, where Im($\varepsilon$) and Re($\varepsilon$) are the imaginary and the real part of the (relative) complex permittivity $\varepsilon$ (reduced to its value of free space), was selected to study a dielectric relaxation spectra in the frequency domain. The latter exhibits (within the Debye approximation) a maximum at a frequency

$$f_{max,\tan\delta} = \sqrt{\varepsilon_s/\varepsilon_\infty}/(2\pi\tau) \quad (4)$$

where $\tau$ the relaxation time (Eq. (1)), $\varepsilon_s$ and $\varepsilon_\infty$ denote the static and high-frequency (relative) permittivity, respectively. These values can be estimated by extrapolating the available broadband complex impedance data to the low and high frequency limit, respectively. Differentiating the natural logarithm of Eq. (4) with respect to pressure at constant temperature, we get:

$$\left(\partial \ln\tau/\partial P\right)_T = \tfrac{1}{2}\left(\partial \ln(\varepsilon_\infty/\varepsilon_s)/\partial P\right)_T - \left(\partial \ln f_{max,\tan\delta}/\partial P\right)_T \quad (5)$$

From Eqs (2) and (5), we get the following formula, which permits the evaluation of the activation volume from the pressure variation of $f_{max,\tan\delta}$ and the pressure variation of $\varepsilon_\infty/\varepsilon_s$:

$$\upsilon^{act} = k_B T\left\{\frac{\gamma}{B} + \frac{1}{2}(\partial\ln(\varepsilon_\infty/\varepsilon_s)/\partial P))_T - (\partial \ln f_{max,\tan\delta}/\partial P)_T\right\} \quad (6)$$

**4. Experimental results and discussion**

A typical plot of tan$\delta$ vs frequency of wetted granodiorite at constant temperature and pressure is presented in Fig. 1, where a relaxation mechanism appears below the kHz region. The latter is sensitive to the values of hydrostatic pressure, as can be seen in Fig. 2: at constant temperature, its maximum shifts toward higher frequency on compression, indicating that the process related with the above mentioned relaxation becomes faster on increasing the pressure. The corresponding isothermal variation of $f_{max,\tan\delta}$ vs is depicted in Fig. 3. A straight line with slope $\left(\partial \ln f_{max,\tan\delta}/\partial P\right)_T = (6.0 \pm 0.9)\text{GPa}^{-1}$ can fit the data points. On the other hand, the pressure dependence of $\varepsilon_\infty$ and $\varepsilon_s$ yields $\left(\partial \ln(\varepsilon_\infty/\varepsilon_s)/\partial P\right)_T \approx 0.5\text{GPa}^{-1}$. However, the extraction of the activation volume through Eq. (6) (within the Debye approximation) presumes an Arrhenius behavior of $\tau(T)$ isobars. The latter was actually verified in the temperature range from 304 K to 363 K; a typical value for the activation enthalpy (Varotsos and Alexopoulos 1978; 1979) at P=292MPa is $0.48 \pm 0.08$ eV. The validity of the Arrhenius law permits subsequently the evaluation of the activation volume by using Eq. (6): $\upsilon^{act} = (-9 \pm 2)\text{cm}^3/\text{mole}$. It is difficult to correlate the observed relaxation mechanism with specific microscopic processes due to the complexity and heterogeneity of partially saturated granodiorite. The relaxation peak can be attributed to the dynamic behavior of various 'dipole' formations, like defect dipole rotation, interfacial polarization, dynamics of water molecules in confined areas or surfaces, or, as an effective result of different cooperating constituents etc. It is worth noticing that the absolute value of $\upsilon^{act}$ reported in the present work is roughly 40% of the molar volume of quartz, which is a dominant component of granodiorite rocks. The relaxation process is therefore probably related with the dynamics of entities of size of the order of the mean atomic (ionic) volume.

It seems that the negative sign in the activation volume for relaxation in granodiorite, which, as mentioned in the second paragraph, was rarely reported for solids, can probably link to the sensitivity phenomenon observed in stations located in granodiorite dykes, as reported in earlier field experiments (Varotsos and Lazaridou (1991)). The present work focuses merely on one parameter that might play role to the sensitivity effect, i.e., the dielectric properties of the (granodiorite), which – according to our (laboratory) results – belongs to the special category of solids where relaxation is monitored by negative activation volume. Actually, the negative activation volume reported here is a pre-requisite for the SES generation. Once the SES has been emitted, in order to be detectable at a site located on Earth's surface (sensitivity) the following may happen as suggested by Varotsos and Alexopoulos (Varotsos and Alexopoulos 1986): the earthquakes usually occur close to pre-existing faults, which are highly conductive paths that may end just below the Earth's surface (for example, see Fig. 25 of Varotsos, Alexopoulos and Lazaridou 1993). Thus, most of the current density of the emitted SES signal follows the neighboring conducting path. If the measuring station is located close to the upper end of this path, the SES electric field is intensified (due to edge effects), thus becoming measurable na hence the site is termed 'sensitive'. In summary, it is currently believed (Varotsos, Alexopoulos and Lazaridou 1993) that the SES sensitivity is associated with electrical inhomogeneities in the Earth's crust.

## 5. Conclusion

The present laboratory results indicate that the dielectric relaxation in wetted granodiorite from Keratea (Greece) is governed by an activation volume of negative sign, which is rarely reported in the literature. The present finding is important for two reasons: (i) the negative activation volume found for the granodiorite may probably link to the sensitivity to SES of Keratea VAN station, and (ii) thereby, supports - in addition to our recent results (Papathanassiou et al (2010)) – the theoretical model, which was suggested as a physical explanation for the generation of transient electric signals long before an earthquake.

Papathanassiou A.N. (2001a), On the polarization mechanisms related to the liquid-solid interaction, J. Phys. Condens. Matter 13, L791-L782 doi: 10.1088/0953-8984/12/26/324

Papathanassiou A.N. and Grammatikakis J. (1996b), Pressure variation of the electrical conductivity of dolomite $CaMg(CO_3)_2$, Phys.Rev.B 53,16247-16251 doi: 10.1103/PhysRevB.53.16247

Papathanassiou A.N. (1998), Effect of hydrostatic pressure on the electrical conductance of polycrystalline magnesite ($MgCO_3$), Phys. Rev. B 58, 4432-4437

Papathanassiou A.N. (2001c), Pressure variation of the conductivity in single crystal calcite, Phys. Stat. Solidi (b) 228, R6-R7

Papathanassiou A.N. (2002), Dependence of the electrical conductivity and the low-frequency dielectric constant upon pressure in porous media containing a small quantity of humidity, Electrochim. Acta 48, 235-239 doi: 10.1016/S0013-4686(02)00619-9

Papathanassiou A.N., Sakellis I. and Grammatikakis J. (2006), Separation of electric charge flow mechanisms in conducting polymer networks under hydrostatic pressure, Appl. Phys. Lett. 89, 222905 doi: 10.1063/1.2768623

Papathanassiou A.N., Sakellis I. and Grammatikakis J. (2007), Migration volume for polaron dielectric relaxation in disordered materials, Appl. Phys. Lett. 91, 202103 doi: 10.1063/1.2812538

Papathanassiou A.N., Sakellis I. and Grammatikakis J. (2010), Negative activation volume for dielectric relaxation in hydrated rocks, Tectonophysics 490, 307-309

Surkov V.V., Uyeda S., Tanaka H. and Hayakawa M. (2002), Fractal properties of medium and seismoelectric phenomena, J. Geodynamics 33, 477-487

Varotsos P. (2007) Comparison of models that interconnect point defect parameters in solids with bulk properties, J. Appl. Phys. 101, 123503

Varotsos P. and Lazaridou M. (1991) Latest aspects of earthquake prediction in Greece based on seismic electric signals, Tectonophysics 188, 321-347

Varotsos P. and Alexopoulos K. (1978) Curvature in conductivity plots of silver halides as consequence of anharmonicity, J. Phys. Chem. Solids, 39, 759-761

Varotsos P. A., Sarlis N. V., Skordas E., Tanaka H.K. and Lazaridou M.S. (2006a), Attempt to distinguish long-range temporal correlations from the statistics of the increments by natural time analysis Phys. Rev. E 74, 021123 doi 10.1103/PhysRevE.74.021123

Varotsos P. A., Sarlis N. V., Skordas E., Tanaka H.K. and Lazaridou M.S. (2006b), Entropy of seismic electric signals: Analysis in natural time under time reversal, Phys. Rev. E 73, 031114 doi 10.1103/PhysRevE.73.031114

Wang Chi-Yuen (1974), Pressure coefficient of compressional wave velocity for a bronzite, J. Geophys. Res. 79, 771-772

Page 12

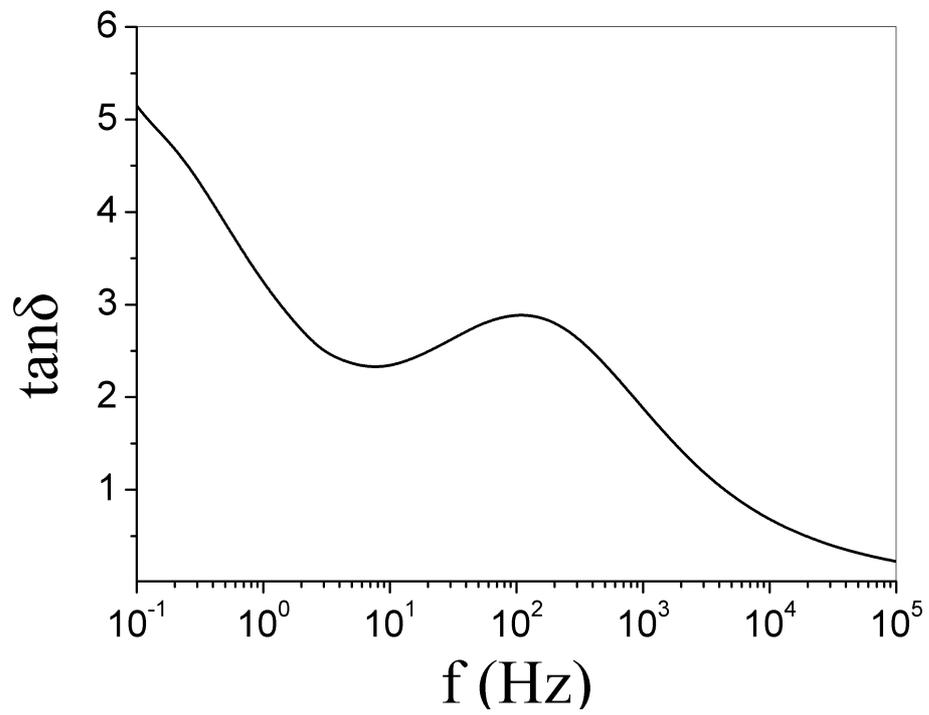

**Figure 1.** A typical plot of tanδ vs frequency of wetted granodiorite recorded at constant temperature (T=333 K) and constant pressure (195 MPa).

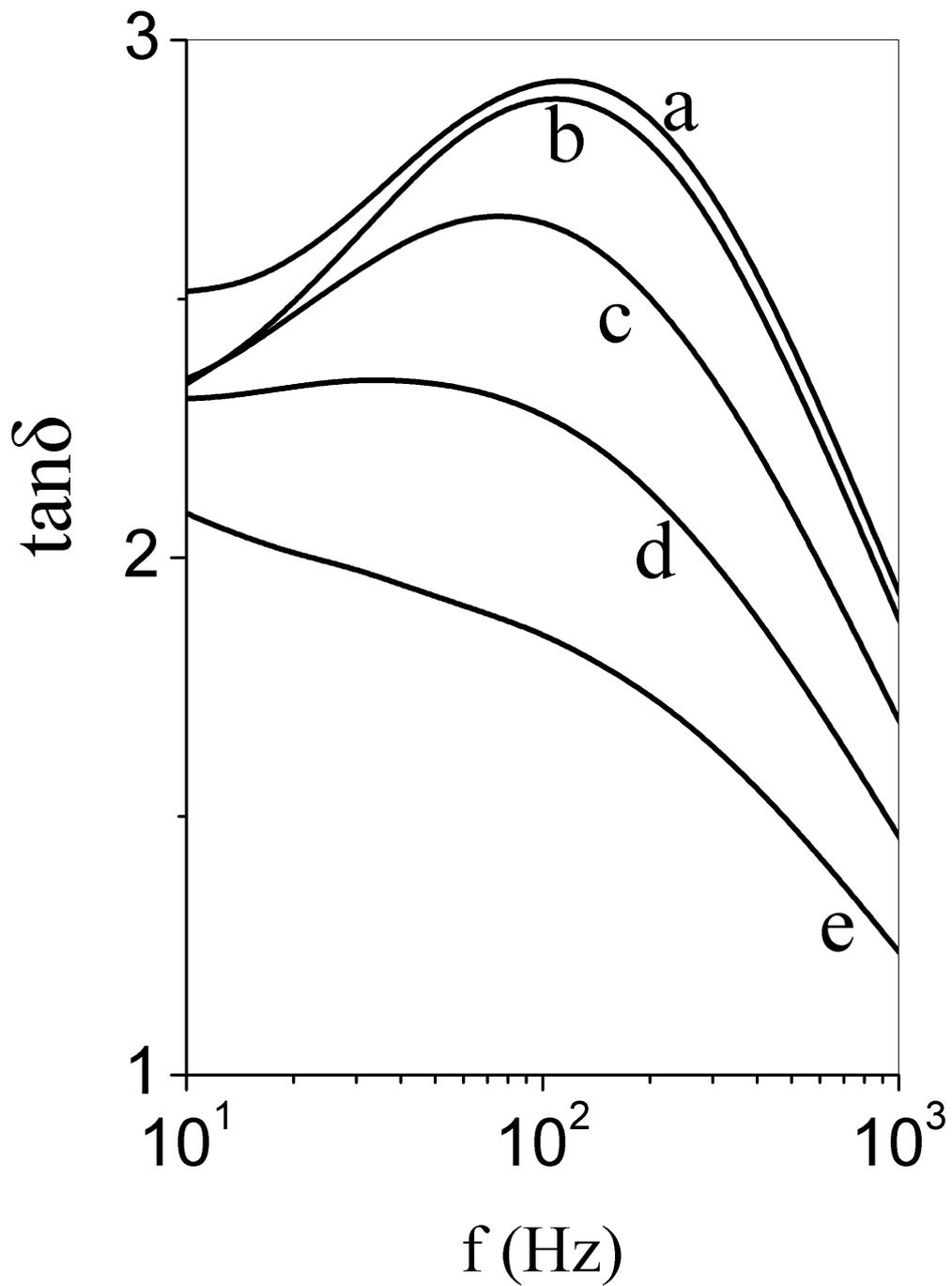

**Figure 2.** Isotherms of tanδ vs frequency of wetted granodiorite recorded at constant temperature (T=333 K) and various pressures: (a): 245 MPa, (b): 195 MPa, (c): 125 MPa, (d): 59 Mpa and (e): 0.1 MPa (ambient pressure).

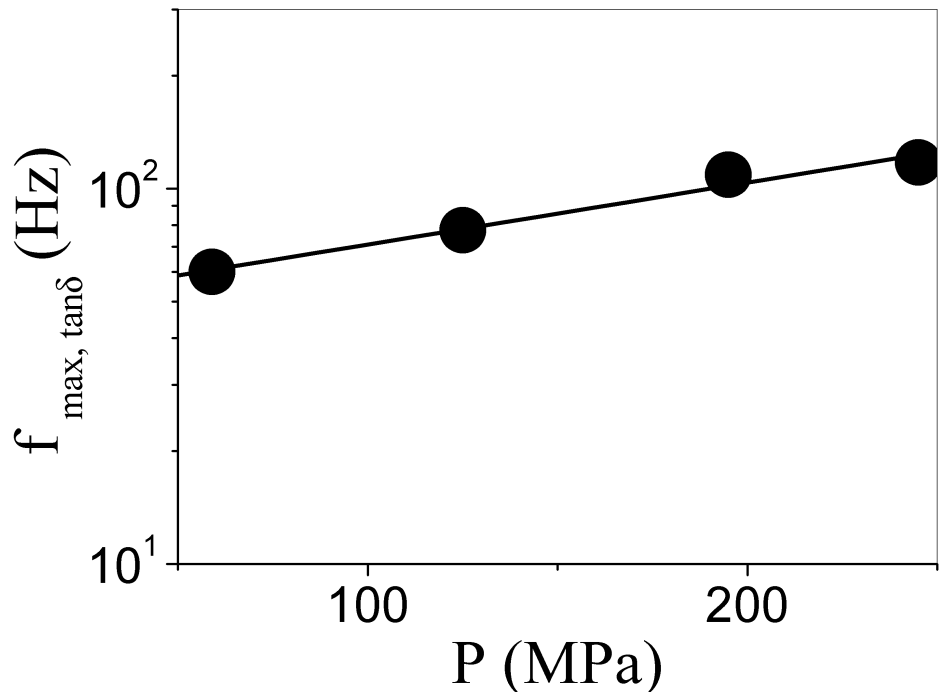

**Figure 3.** Pressure dependence of $f_{max,\,\tan\delta}$ at T=333 K. The straight line is a best fit to the data points.